\documentclass[a4paper,12pt]{article}
\usepackage{latexsym}
\usepackage{rotating}
\usepackage{amsmath}
\author{Wen-Xiu~Ma\footnote{Email: mawx@math.cityu.edu.hk} \\
\small {Department of Mathematics, City University of Hong Kong,
Kowloon, Hong Kong, P. R. China }\vspace{1mm}\\
Si-Ming Zhu\footnote{Email: stszsm@zsulink.zsu.edu.cn}\\
{\small Department of Mathematics,
Zhongshan University, Guangzhou 510275, P. R. China}}
\title{Non-symmetry constraints of the AKNS system yielding 
integrable Hamiltonian systems 
}
\date{\nonumber}

\begin{document}
\maketitle

%%%%%%%%%%%%%%%%%%%%%%%%%%%%%%%%%%%%%%%%%%%%%%%%%%%%%%%%%%%%%%%%%%%%%%%%
\def\be {\begin{equation}}
\def\ee{\end{equation}}
\def\bea{\begin{eqnarray}}
\def\eea{\end{eqnarray}}
\def\ba{\begin{array}}
\def\ea{\end{array}}
\def\la {\lambda}
\def \part {\partial}
\def \al {\alpha}
\def \de {\delta}

\newcommand{\R}{\mbox{\rm I \hspace{-0.9em} R}}

\renewcommand{\theequation}{\thesection.\arabic{equation}}

%%%%%%%%%%%%%%%%%%%%%%%%%%%%%%%%%%%%%%%%%%%%%%%%%%%%%%%%%%%%%%%%%%%%%%%%

\begin{abstract}

This paper aims to show that 
there exist non-symmetry constraints which yield
integrable Hamiltonian systems
through nonlinearization of spectral problems
of soliton systems, like symmetry constraints.
Taking the AKNS spectral problem as an illustrative example,
a class of such non-symmetry constraints is introduced 
for the AKNS system,
along with two-dimensional integrable Hamiltonian systems 
generated from the AKNS spectral problem. 

\end{abstract}

\section{Introduction}
\setcounter{equation}{0}

The nonlinearization process yields integrable Hamiltonian systems
from spectral problems of soliton systems \cite{Cao-SC1990}-\cite{Ma-book1990}.
Much excitement in the study of nonlinearization
comes from a kind of specific symmetry constraints \cite{MaS-PLA1994,Ma-JPSJ1995}.
%:\[K_0=J\psi ^T\frac {\partial U}{\partial \lambda }\psi,  \]
%where $K_0$ is a symmetry and $J$ is a Hamiltonian operator   
It is due to symmetry constraints that the nonlinearization technique
is so powerful in generating integrable Hamiltonian systems 
\cite{MaFO-PA1996,MaDZL-NCB1996}.
Usually, taking non-symmetry constraints 
leads to spectral problems to non-Hamiltonian systems, which 
are difficult to be handled. However, there appears  
a natural question of whether there exist any non-symmetry constraints 
which can still force spectral problems of soliton systems 
to be integrable Hamiltonian systems. 
The answer is yes. 
%Even if we do not employ symmetry constraints, 
This paper aims to show that it is possible to generate 
integrable Hamiltonian systems from spectral problems of soliton systems by employing non-symmetry constraints of soliton systems.

In the following section, we 
take the AKNS spectral problem as an illustrative example and 
recall some known results related to symmetry constraints 
of the AKNS system for reference.
Then in Section \ref{Section3}, we move on to discuss
non-symmetry constraints of the AKNS system. 
A class of non-symmetry constraints
is introduced for the AKNS system,
which generates two-dimensional integrable Hamiltonian systems 
from the AKNS spectral problem. 
Finally in Section \ref{Section4}, a comparison between our integrable systems and other integrable systems and some concluding
remarks are given.

\section{AKNS system and symmetry constraints}
\setcounter{equation}{0}

Let us %shed right on the idea 
take the AKNS spectral problem \cite{AblowitzKNS-SAM1974}
\begin{equation} \phi_x=U\phi,\ \phi=
\left( \ba {c}  \phi _1  \vspace{2mm}\\
\phi _2 \ea  \right),\ U=U(u,\la )=
\left( \ba {cc}  -\la & q   \vspace{2mm}\\
r & \la  \ea  \right),\ u=
\left( \ba {c}  q  \vspace{2mm}\\
r \ea  \right),
 \label{spofAKNS}
\end{equation}
with $\la $ being a spectral parameter, as an illustrative example.
Its adjoint spectral problem %of the AKNS spectral problem (\ref{spofAKNS})
reads as  
\begin{equation} \psi _x=-U^T\psi ,\  \psi =
 \left( \ba {l}  \psi _1  \vspace{2mm}\\
\psi _2 \ea  \right),\ \textrm{i.e.}\ 
\left( \ba {l}  \psi _1  \vspace{2mm}\\
\psi _2 \ea  \right)_x =
\left( \ba {cc}  \la & -r   \vspace{2mm}\\
-q & -\la  \ea  \right)
\left( \ba {l}  \psi _1  \vspace{2mm}\\
\psi _2 \ea  \right),\label{aspofAKNS}
\end{equation}
where $T$ denotes the transpose operation of matrices.
The spectral problem (\ref{spofAKNS}) or its adjoint 
spectral problem (\ref{aspofAKNS}) yields the AKNS hierarchy 
\begin{equation} u_t=
\left( \ba {l}  q \vspace{2mm}\\
r\ea  \right)_t=K_n= \Phi ^n 
\left( \ba {c}  -2q  \vspace{2mm}\\
2r \ea  \right)=JG_n=J\frac {\delta {\tilde H}_n}{\delta u},
\  n\ge 0,
\label{AKNShierarchy} \end{equation}
where the Hamiltonian operator $J$, the recursion operator $\Phi$, and
the Hamiltonian functionals ${\tilde H}_n$ are defined by 
\begin{eqnarray}&& J=\left( \ba {cc}  0 & -2   \vspace{2mm}\\
2 &0 \ea  \right),\ \Phi =
\left( \ba {cc} 
-\frac 12 \part + q\part ^{-1}r &
  q\part ^{-1} q
\vspace{2mm}\\
 - r\part ^{-1} r &
 \frac 12 \part -  r\part ^{-1} q
 \ea  \right),\\&&  
{\tilde H}_n=\int H_n\,dx ,\ 
H_n=\int _0^1<G_n(\la u),u>\,d\la ,\ n\ge 0,
\end{eqnarray}
where $<\cdot,\cdot>$ denotes the standard inner product of $\R ^2$.
The first nonlinear integrable system in this 
soliton hierarchy is the AKNS system of nonlinear Schr\"odinger equations
%The first nonlinear integrable system in this AKNS hierarchy reads as
\begin{equation}   u_t=
\left( \ba {l}  q \vspace{2mm}\\
 r\ea  \right)_t= K_2=
\left( \ba {l} -\frac 12 q_{xx}+q^2r  \vspace{2mm}\\
 \frac 12 r_{xx}-qr^2 \ea  \right).\label{firstAKNSsystem}
\end{equation}
This AKNS system has its associated spectral problem
\begin{equation} \phi_t=V^{(2)}\phi ,\ 
V^{(2)}=V^{(2)}(u,\lambda )=
 \left( \ba {cc}  -\la ^2+\frac 12 qr & \la q -\frac 12 q_x   \vspace{2mm}\\
\la r +\frac 12 r_x & \la ^2-\frac 12 qr  \ea  \right),
 \label{tspoffirstAKNS}
\end{equation}
which implies that (\ref{firstAKNSsystem}) is equivalent to 
a zero curvature equation $U_t-V_x^{(2)}+[U,V^{(2)}]=0$.
The AKNS system (\ref{firstAKNSsystem}) also has
a tri-Hamiltonian structure 
\begin{equation} 
u_t=K_2=J_0\frac {\delta {\tilde H}_2}{\delta u}=J_1\frac {\delta {\tilde H}_1}{\delta u}
=J_2\frac {\delta {\tilde H}_0}{\delta u}, \end{equation}
where the Hamiltonian operators $J_i, \ 0\le i\le 2,$
and the Hamiltonian functionals ${\tilde H}_i,\ 0\le i\le 2,$ 
are given by 
\begin{eqnarray}&&
J_0=J=\left( \ba {cc} 0&-2 \vspace{2mm}\\2 &0\ea \right ),\ 
 J_1=\Phi J=\left( \ba {cc} 
 2 q\part ^{-1} q & \part - 2q\part ^{-1}r \vspace{2mm}\\ 
\part - 2 r\part ^{-1} q & 2 r\part ^{-1} r  \ea  \right),\\&& 
J_2=\Phi J_1=
\left ( \ba {cc}  q\part ^{-1}q\part - \part q \part ^{-1}q
&-\frac 12 
 \part ^2 + q\part ^{-1}r\part  + \part q\part ^{-1}r
 \vspace{2mm}\\
\frac 12 \part ^2 - \part r\part ^{-1}q - r\part ^{-1}q\part
&  \part r\part ^{-1}r - r \part ^{-1}r\part 
\ea \right ) ,\\&&
{\tilde H}_0=\int qr\,dx,\ {\tilde H}_1=\frac 14 \int (qr_x- q_xr)\,dx,\ 
{\tilde H}_2= \frac 18 \int (qr_{xx}+q_{xx}r-2 q^2r^2)\,dx .
\qquad \quad
\end{eqnarray}
A proof that $J_0+\alpha J_1+\beta J_2$ is Hamiltonian for all
$\alpha $ and $\beta $ can be found in \cite{MaZ-JMP1999}.
%The AKNS system (\ref{firstAKNSsystem}) is a main object in our discussion. 

It is known \cite{MaS-PLA1994} that 
the spectral problem (\ref{spofAKNS}) 
and the adjoint spectral problem (\ref{aspofAKNS}) 
become a finite-dimensional integrable Hamiltonian system 
\begin{equation} \phi _{ix}=-\frac {\part H}{\part \psi_i},\ 
\psi _{ix}=\frac {\part H}{\part \phi_i},\  
 H=\la \phi_2\psi _2-\la \psi _1\psi _1 +\phi_1\phi_2\psi_1\psi_2, \  i=1,2,
\label{cfofAKNS}\end{equation} 
when we employ a symmetry constraint
\begin{equation} K_0= E J{\delta \la }/{\delta u}=
J\psi ^T\frac {\partial U}{\partial u }\psi =
J\left(\ba {c} \phi_1\psi_2 \vspace{2mm}\\ 
\phi_2\psi_1 \ea \right),  \label{originalformofsymmetryconstraintofAKNS}\end{equation}
with $E$ being a normalized constant,
which leads to two constraints on the potentials 
\begin{equation}
q=\phi_1 \psi _2,\ r= \phi_2\psi _1.\label{symmetryconstraintsofAKNS}
\end{equation}
Note $J\frac {\delta \lambda }{\delta u} $ is a symmetry of the AKNS 
system (\ref{firstAKNSsystem}) 
due to $\la _t=0$. Actually it can directly be shown that
$J( \phi_1\psi_2,\phi_2\psi_1 )^T$ is a symmetry of the AKNS system (\ref{firstAKNSsystem}),
that is to say,
\[
\sigma =\left(\ba {c} \sigma _1\vspace{2mm}\\ \sigma_2 \ea \right)=
\left(\ba {c} -2\phi_1\psi_2\vspace{2mm}\\ 2\phi_2\psi_1\ea \right)
\] 
satisfies the linearized system of the AKNS system (\ref{firstAKNSsystem}):
\begin{equation}
 \sigma _{1t}= -\frac 12\sigma _{1xx}+2qr \sigma _1+q^2\sigma _2,
\ \sigma _{2t}= \frac 12\sigma _{2xx}-r^2 \sigma _1-2qr \sigma _2,
\label{linearizedsystemoffirstAKNS}
\end{equation}
when $\phi$ and $\psi$ satisfy two systems of (\ref{spofAKNS})
and (\ref{aspofAKNS}) and evolve according to 
\begin{equation}
 \phi _{t}=V^{(2)}\phi =V^{(2)}(u,\la )\phi , \ \psi_{t}=-(V^{(2)})^T\psi =
-(V^{(2)}(u,\la ))^T\psi , \label{tspaspoffirstAKNS}
\end{equation}
with $V^{(2)}$ being defined by (\ref{tspoffirstAKNS}).
Therefore, (\ref{originalformofsymmetryconstraintofAKNS}) is  
a symmetry constraint indeed, because $K_0$ on the left side of
(\ref{originalformofsymmetryconstraintofAKNS}) 
and $J{\delta \la }/{\delta u}$
on the right side of (\ref{originalformofsymmetryconstraintofAKNS}) 
are all symmetries of the AKNS system (\ref{firstAKNSsystem}).
Moreover, we can show that (\ref{originalformofsymmetryconstraintofAKNS})
is a symmetry constraint of each system in the AKNS hierarchy 
(\ref{AKNShierarchy}). 

\section{Non-symmetry constraints yielding integrable Hamiltonian systems}
\setcounter{equation}{0}
\label{Section3}

In what follows, we are going to present a class of 
non-symmetry constraints that still yield integrable Hamiltonian 
systems through nonlinearization of the spectral problem (\ref{spofAKNS}) 
and the adjoint spectral problem (\ref{aspofAKNS}). 
Let us first assume that 
two constraints between the potentials and the eigenfunctions and 
adjoint eigenfunctions are defined by 
\begin{equation} q=q(\phi_1,\phi_2,\psi_1,\psi_2),\ 
r= r(\phi_1,\phi_2,\psi_1,\psi_2). 
\label{gconstraintsofAKNS}
\end{equation}
Notice that 
a vector-valued function $f:\R ^n\to \R^n$, i.e. 
\[f=(f_1(x_1,\cdots ,x_n),\cdots ,f_n(x_1,\cdots,x_n)),\ (x_1,\cdots ,x_n)
\in \R^n,\]
is a gradient $f=\textrm{grad}\,h$ of some function 
$h:\R^n\to \R$
if and only if 
\[ \frac {\part f_i}{\part x_j}= 
 \frac {\part f_j}{\part x_i},\ 1\le i<j\le n. \]
It follows that 
the spectral problem (\ref{spofAKNS}) 
and the adjoint spectral problem (\ref{aspofAKNS}) 
is a Hamiltonian system under the constraints (\ref{gconstraintsofAKNS})
and the symplectic form 
\begin{equation} \omega ^2
=d \phi_1\wedge d\psi_1 + d \phi_2\wedge d\psi_2 ,\end{equation} 
if and only if 
\begin{equation} \left\{
%{\displaymath 
\ba {l} 
{\displaystyle 
\frac {\part q}{\part \phi_1}\phi_2=
\frac {\part r}{\part \psi_1}\psi_2, \ 
\frac {\part q}{\part \psi_2}\psi_1=
\frac {\part r}{\part \phi_2}\phi_1, }
\vspace{2mm} \\
{\displaystyle 
 \frac {\part q}{\part \phi_1}\psi_1=
\frac {\part r}{\part \phi_2}\psi_2, \ 
\frac {\part q}{\part \psi_2}\phi_2=
\frac {\part r}{\part \psi_1}\phi_1, }\vspace{2mm} 
\\
 {\displaystyle 
\frac {\part q}{\part \phi_2}\phi_2=
\frac {\part q}{\part \psi_1}\psi_1, \ 
\frac {\part r}{\part \phi_1}\phi_1=
\frac {\part r}{\part \psi_2}\psi_2. }
\ea  \right. 
\label{keyequation} \end{equation} 
Our case has $ n=4$, and the functions $f_i$ and the variables $x_i$ are 
chosen to be 
\begin{eqnarray} &  & f_1=\lambda \psi_1-r\psi_2\,(=\psi_{1x}), \ 
f_2= -q\psi_1-\lambda \psi_2\,(=\psi_{2x}),  \nonumber  \\ &  &
f_3=\lambda  \phi_1-q\phi_2 \,(=-\phi_{1x}), \ f_4= -r\phi_1-\lambda 
\phi _2\,(= -\phi _{2x}), 
  \nonumber 
\end{eqnarray}
and 
\begin{equation}
(x_1,x_2,x_3,x_4)=(\phi _1,\phi_2,\psi_1,\psi_2).\nonumber
\end{equation}
We point out that from (\ref{keyequation}) we can further obtain 
two conditions similar to the last two conditions in (\ref{keyequation}):
\[ \frac {\part q}{\part \phi_1}\phi_1=
\frac {\part q}{\part \psi_2}\psi_2, \ 
\frac {\part r}{\part \phi_2}\phi_2=
\frac {\part r}{\part \psi_1}\psi_1. \]

Now the construction of the constraints (\ref{gconstraintsofAKNS})
yielding Hamiltonian systems
becomes the problem of finding solutions to 
the system of differential equations (\ref{keyequation}).
Fortunately by inspection, a solution to the system (\ref{keyequation}) is found to be 
\begin{equation} 
q=\al \phi _1\psi _2 +g_1(\phi_2\psi_1),\ 
r=\al \phi _2\psi _1 +g_2 (\phi_1 \psi_2),
\label{specialconstraints}\end{equation}
where $\al $ is an arbitrary constant, and $g_1 ,g_2 :\R \to \R$ 
are two arbitrary functions.
Changing the constraints (\ref{specialconstraints}) 
into the following form
\begin{equation} K_0=\left( \ba {c} -2q\vspace{2mm}\\ 2r\ea \right)=
\left( \ba {c} -2(\al \phi _1\psi _2 +g_1(\phi_2\psi_1))
\vspace{2mm}\\ 2(\al \phi _2\psi _1 +g_2( \phi_1 \psi_2))
\ea \right) =
\al \left( \ba {c} -2 \phi _1\psi _2 \vspace{2mm}\\ 2 \phi _2\psi _1 \ea \right)
+\left( \ba {c} -2g_1(\phi_2\psi_1)
\vspace{2mm}\\ 2 g_2( \phi_1 \psi_2)\ea \right)
,
  \label{generalconstraintofAKNS}\end{equation} 
it is not difficult to see that 
the constraints (\ref{specialconstraints}) with $g_1^2+g_2^2 \ne  0$
are not generated from symmetry constraints of the AKNS system (\ref{firstAKNSsystem}),
because the vector fields $K_0$ and $(-2\phi_1\psi_2,2\phi_2\psi_1)^T$
are symmetries but all nonzero vector fields 
\begin{equation}
\left(\begin{array}{l}\sigma _1\vspace{2mm}\\\sigma_2
\end{array}\right)=\left(\begin{array}{c}
-2g_1(\phi_2\psi_1)\vspace{2mm}\\
2g_2(\phi_1 \psi_2)\end{array}\right)
\end{equation}
are not symmetries of the AKNS system (\ref{firstAKNSsystem}). 
This can be shown by observing the terms involving $q^2$ and $r^2$ 
in the linearized system (\ref{linearizedsystemoffirstAKNS}).
Keeping (\ref{spofAKNS}), (\ref{aspofAKNS}) and (\ref{tspaspoffirstAKNS})
in mind, we find that $\sigma _{1t},\,-1/2\sigma_{1xx}$ and $2qr\sigma _1$
%\[\ba {l} 
% \sigma _{1t}= , \vspace{2mm}\\
%-1/2\sigma_{1xx}= , \vspace{2mm}\\ 
%2qr\sigma _1= \ea  \] 
do not contain any term involving $q^2$ and thus the first equation of the linearized system
(\ref{linearizedsystemoffirstAKNS}) requires $g_2=0$. Similarly, we can find that the second equation of the linearized system
(\ref{linearizedsystemoffirstAKNS}) requires $g_1=0$. 
This will contradict our assumption of the nonzero condition on 
$(\sigma_1,\sigma_2)^T$: $g_1^2+g_2^2\ne  0$. Therefore,  
$(\sigma_1,\sigma_2)^T=(-2g_1(\phi_2\psi_1),2g_2(\phi_1 \psi_2))^T$ 
with $g_1^2+g_2^2\ne 0$ are not symmetries of 
the AKNS system (\ref{firstAKNSsystem}), and so the constraints (\ref{generalconstraintofAKNS}) with $g_1^2+g_2^2\ne 0$ 
are not symmetry constraints of (\ref{firstAKNSsystem}),
because $K_0-2\alpha (-\phi_1\psi_2,\phi_2\psi _1)^T$ is a symmetry 
of (\ref{firstAKNSsystem}). Moreover, we believe
that (\ref{generalconstraintofAKNS}) is not a symmetry constraint
of the other systems in the AKNS hierarchy (\ref{AKNShierarchy}), 
either.

Let us now take the constraints (\ref{specialconstraints}), and then
the spectral problem (\ref{spofAKNS}) 
and the adjoint spectral problem (\ref{aspofAKNS}) 
are nonlinearized into a Hamiltonian system 
\begin{equation}
\phi _{ix}=-\frac {\part H(g_1,g_2)}{\part \psi_i},\ 
\psi _{ix}=\frac {\part H(g_1,g_2)}{\part \phi_i},\ i=1,2,
\label{Hamiltoniansystemundersc}
\end{equation}
with the Hamiltonian function 
\begin{equation}
H(g_1,g_2)=\la (\phi_1\psi _1-\phi_2\psi_2)-\al \phi_1\phi_2\psi_1\psi_2-
h_1(\phi_2\psi_1)-h_2(\phi_1\psi_2)
%\frac {\beta }{m+1}\phi_2^{m+1}\psi_1^{m+1}- 
%\frac {\beta }{m+1}\phi_1^{m+1}\psi_2^{m+1}
, \label{Hamiltonianfunctionundersc}
\end{equation} 
where $h_1,h_2:\R\to \R$ are two anti-derivative functions of $g_1,g_2$, respectively.
This Hamiltonian system has a second integral of motion
\begin{equation} F = \phi_1\psi_1+\phi_2\psi_2. \label{F}\end{equation}
It can be generated as follows \cite{MaS-PLA1994,Palais-BAMS1997}
\[ F=\textrm{tr}\bar V ,\ \bar V=\phi \psi ^T= 
\left( \ba {c}  \phi _1  \vspace{2mm}\\
\phi _2 \ea  \right)(  \psi _1 ,\psi _2 ) =
\left( \ba {cc}  \phi _1\psi_1&\phi_1\psi_2  \vspace{2mm}\\
\phi _2\phi_1&\phi_2\psi_2 \ea  \right)
.\]
Since we have $\bar V_x=[U,\bar V]$ provided that (\ref{spofAKNS}) and 
(\ref{aspofAKNS}) hold, we can compute that 
\[ F_x=\textrm{tr}(\bar V_x)= \textrm{tr}[U,\bar V]=0.\]
This means that the function $F$ is an integral of motion
of the spectral problem (\ref{spofAKNS}) 
and the adjoint spectral problem (\ref{aspofAKNS}) 
with any potentials $q$ and $r$, and hence $F$ is also 
an integral of motion the Hamiltonian system defined by 
(\ref{Hamiltoniansystemundersc}) 
and (\ref{Hamiltonianfunctionundersc}), where $q$ and $r$ are the
special functions defined by (\ref{specialconstraints}).

Two integrals of motion $H(g_1,g_2)$ and $F$ are functionally independent 
and of course they commute with each other, i.e.  the Poisson bracket of $H(g_1,g_2)$ and $F$
is equal to zero, 
\[ \{H(g_1,g_2),F\}=<\frac {\part H(g_1,g_2)}{\part \psi},
\frac {\part F}{\part \phi}>-<\frac {\part H(g_1,g_2)}{\part \phi},
\frac {\part F}{\part \psi}>=0, \]
since $F$ is an integral of motion of the Hamiltonian system (\ref{Hamiltoniansystemundersc}).
These two properties guarantee that the Hamiltonian system defined by 
(\ref{Hamiltoniansystemundersc}) and (\ref{Hamiltonianfunctionundersc}) 
is Liouville integrable \cite{Arnold-book1989}. 
Therefore the spectral problem (\ref{spofAKNS}) 
and the adjoint spectral problem (\ref{aspofAKNS}) 
are nonlinearized into an integrable Hamiltonian system under the constraints (\ref{specialconstraints}), which are not of symmetry type
when $g_1^2+g_2^2\ne 0$. 
%The whole process of nonlinearization provides us with 

The class of integrable Hamiltonian systems generated above has two 
degrees of freedom. They contain the Hamiltonian system (\ref{cfofAKNS}) 
under a symmetry constraint if we take $\al =1$ and $g_1=g_2=0$. 
Another interesting integrable Hamiltonian system is associated 
with the case of $\al =0$ and $g_1(y)=g_2(y)=y$:
\begin{equation}
\phi_{1x}=-\la \phi_1+\phi_2^2\psi_1,\ \phi_{2x}=\phi_1^2\psi_2+\la \phi _2,\ 
\psi_{1x}=\la \psi_1-\phi_1\psi_2^2,\ \psi_{2x}=-\phi_2\psi_1^2-\la \psi_2.\label{1stspecial2DIHS}
\end{equation}
This system interchanges 
two constraints on the potentials $q$ and $r$ in the symmetry case
(\ref{symmetryconstraintsofAKNS})
but it corresponds to a non-symmetry case. 
A more general example can be presented 
by choosing 
\begin{equation}
g_1(y)=h_1(y)=\beta _1y^m+\gamma _1 \textrm{e}^{y} ,\ 
g_2(y)=h_2(y)=\beta _2y^n+\gamma _2 \textrm{e}^{y} ,
\end{equation}
where $\beta _i$ and $\gamma _i$ are arbitrary constants and 
$m,n$ are non-negative integers. 
The resulting integrable system reads as
\begin{equation}
\left \{ \ba {l} 
\phi_{1x} =-\la \phi_1 + \al \phi_1\phi_2\psi _2 +\beta _1 \phi_2^{m+1}\psi_{1}^m +\gamma _1 \phi _2\textrm{e}^{\phi_2\psi _1}, \vspace{2mm}\\ 
\phi_{2x} =\la \phi_2 + \al \phi_1\phi_2\psi _1 +\beta _2 \phi_1^{n+1}\psi_{2}^n +\gamma _2 \phi _1\textrm{e}^{\phi_1\psi _2}, \vspace{2mm}\\ 
\psi_{1x} =\la \phi_1 - \al \phi_2\psi_1\psi _2 -\beta _2 \phi_1^{n}\psi_{2}^{n+1} -\gamma _2 \psi _2\textrm{e}^{\phi_1\psi _2}, \vspace{2mm}\\ 
\psi_{2x} =-\la \psi_2 - \al \phi_1\psi_1\psi _2 -\beta _1 \phi_2^{m}\psi_{1}^{m+1} -\gamma _1 \psi _1\textrm{e}^{\phi_2\psi _1}, 
\ea \right. 
\label{2ndspecial2DIHS}
\end{equation}
which can be put into the following Hamiltonian system
\begin{equation}
\phi_{ix}=-\frac {\part H_s(h_1,h_2)}{\part \psi _i},\ \psi_{ix}=
\frac {\part H_s(h_1,h_2)}{\part \phi_i},\ i=1,2
\end{equation}
with the Hamiltonian function
\begin{equation}\ba {l} 
H_s(h_1,h_2)=\la (\phi_1\psi_1-\phi_2\psi_2) -\al \phi_1\phi_2\psi_1\psi_2
-\frac {\beta _1}{m+1}(\phi_2\psi_1)^{m+1} \vspace{2mm} \\
\qquad \qquad \quad \  -
\frac {\beta _2}{n+1}(\phi_1\psi_2)^{n+1}
-\gamma _1\textrm{e}^{\phi_2\psi_1} - \gamma _2\textrm{e}^{\phi_1\psi_2}.
\label{generalexample}\ea 
\end{equation}

\section{Concluding remarks}
\setcounter{equation}{0}
\label{Section4}

Our integrable systems above are just a class of 
two-dimensional integrable Hamiltonian systems. 
Each term of the Hamiltonian functions defined by 
%(\ref{Hamiltoniansystemundersc}) and 
(\ref{Hamiltonianfunctionundersc}) 
mixes two kinds of the variables $\phi_1,\phi_2$ and the variables 
$\psi_1,\psi_2$, which shows a different feature from
some well-known dynamical systems such as the St\"ackel systems 
\cite{Perelomov-book1990}, the
many-body systems of interacting particles \cite{Perelomov-book1990,Calogero-LNC1975,Moser-AM1975}, the Toda lattice \cite{Toda-book1989}.
Some specific interesting cases of two-dimensional integrable Hamiltonian  
with polynomial energy were also analyzed
(see, for example,
\cite{GrammaticosDR-JMP1983,Hietarinta-PLA1983,WojciechowskaW-PLA1984}).
On the other hand, a general Hamiltonian function $H=H(\phi _1,\phi_2,\psi_1,\psi_2)$ that commutes with $F$ defined by (\ref{F}) can be found by solving a specific differential equation
\[ \phi_1\frac {\partial H}{\partial \phi_1}+ \phi_2\frac {\partial H}{\partial \phi_2}
=\psi_1\frac {\partial H}{\partial \psi_1}+ \psi_2\frac {\partial H}{\partial \psi_2}.\] 
But the resulting Hamiltonian systems may not be associated with spectral problems of soliton systems. 

We emphasize that the paper aims to provide an example 
that non-symmetry constraints generate integrable Hamiltonian systems 
from spectral problems of soliton systems. Our result shows that
the constraints 
between potentials and eigenfunctions and/or adjoint eigenfunctions
yielding integrable Hamiltonian systems can be both of symmetry type and of non-symmetry type. 
However, non-symmetry constraints are not so powerful as symmetry constraints in generating integrable Hamiltonian systems.
In the AKNS case, we don't think that time parts of Lax pairs,
for example, the system (\ref{tspaspoffirstAKNS}),
can be transformed into integrable Hamiltonian systems
under the non-symmetry constraint (\ref{generalconstraintofAKNS}),
although (\ref{specialconstraints}) provides a B\"acklund transformation 
of the integrable AKNS system (\ref{firstAKNSsystem}).  
It is also interesting to solve the above two-dimensional integrable
Hamiltonian systems, especially (\ref{1stspecial2DIHS}) and (\ref{2ndspecial2DIHS}),
and to extend them to many-body integrable Hamiltonian systems. 

%\newpage
\vskip 3mm
\noindent{\bf Acknowledgments:} This work was in part supported by
a strategic research grant from the City University of Hong Kong (Project No. 7000803) and a competitive earmarked research grant from the Research 
Grants Council of Hong Kong (Project No. 9040395),
and the National Natural Science Foundation of China. 
One of the authors (Ma) is also grateful to Prof. Yishen Li for 
valuable discussion.

%\vskip 2mm

\end{document}